\documentstyle[12pt]{article}
\newlength{\extraspace}
\newlength{\extraspaces}
\newcommand{\be}{\begin{equation}
\addtolength{\abovedisplayskip}{\extraspaces}
\addtolength{\belowdisplayskip}{\extraspaces}
\addtolength{\abovedisplayshortskip}{\extraspace}
\addtolength{\belowdisplayshortskip}{\extraspace}}
\newcommand{\ee}{\end{equation}}

\newcommand{\bq}{\begin{eqnarray}
\addtolength{\abovedisplayskip}{\extraspaces}
\addtolength{\belowdisplayskip}{\extraspaces}
\addtolength{\abovedisplayshortskip}{\extraspace}
\addtolength{\belowdisplayshortskip}{\extraspace}}
\newcommand{\eq}{\end{eqnarray}}
\newcommand{\ra}{\rightarrow}

\begin{document}

\addtolength{\baselineskip}{.8mm}

\thispagestyle{empty}

\begin{flushright}
{\sc PUPT}-1401\\
hep-th@xxx/9305291 \\
 May  1993
\end{flushright}
\vspace{.3cm}

\begin{center}
{\Large  Two-Dimensional Field Theory
 Description of  Disoriented Chiral Condensate.}\\
\vspace{0.4in}
{\large Ian I. Kogan}
\footnote{ On  leave of absence
from ITEP,
 B.Cheremyshkinskaya 25,  Moscow, 117259,     Russia.} \\
\vspace{0.2in}
{\it  Physics Department, Princeton  University \\
 Princeton, NJ 08544
 USA} \\
\vspace{0.7in}
{\sc  Abstract} \\
\end{center}

\noindent
We consider the effective $1+1$-dimensional chiral theory
 describing fluctuations of the order parameter
 of the  Disoriented Chiral Condensate (DCC)  which
 can be formed in the central rapidity region
 in a relativistic nucleus-nucleus or nucleon-nucleon collisions
 at high energy.
   Using  $1+1$-dimensional
 reduction   of QCD at high energies and assuming
   spin polarization of DDC one can find the
 Wess-Zumino-Novikov-Witten  (WZNW) model at level $k=3$ as
   the effective  chiral theory  for  the one-dimensional
 DDC.  Some possible phenomenological consequences are briefly
 discussed.


\vfill

\newpage

\renewcommand{\footnotesize}{\small}

\noindent

There exists some connection between high-energy
scattering in quantum chromodynamics (QCD)  and effective $1+1$ dimensional
 field theory \cite{lipatov}. It  was  shown  recently
 in a very  elegant way \cite{verlinde} how to derive  the effective
 $1+1$- dimensional theory describing the  high energy intraction
 between two quarks in the limit $s >> t >> \Lambda_{QCD}$.
 To get the two-dimensional picture  one can  split four coordinates
 into  two longitudinal
coordinates $x^{\alpha}$ and two transverse coordinates $x^{i}$
\bq
x^{\alpha}  = (x^+,x^-), \;\;\;\;
x^i  =  (x, y)
\eq
with $x^\pm = t \pm z$, and  then
 perform the rescaling of the longitudinal coordinates
  \bq
 x^{\pm} \rightarrow \lambda x^{\pm}
\eq
with $\lambda \sim 1/\sqrt{s} \rightarrow 0$. In this limit QCD Lagrangian
 will reduce to some effective $1+1$-dimensional Lagrangian ( for more details
 see paper\cite{verlinde}).

The aim of this paper is to apply similar ideas to another
 high-energy process - formation of the Disoriented
 Chiral Condensate (DCC) \cite{dcc}  in a relativistic nuclear collision.
     It is well known that QCD  Lagrangian is  invariant
 (approximately iif nonzero masses for the light $N_{f}$  quarks are
 taken into account)
 under global chiral $SU(N_{f})_{L} \times SU(N_{f})_{R}$, where
 $N_{f}$ is the number of the light flavours. This symmetry is
 spontaneously broken down to vector $SU(N_{f})_{V}$ which
 leads to $N_{f}^{2} - 1$ (quasi)goldstone bosons - pions
 (if $N_{f} = 2$) or pions, kaons and $\eta$ meson (if $N_{f} = 3$).
 The order parameter for this breaking is the vacuum expectation
 value of  quark condensate
 $< \bar{\psi}\psi>$. However one can imagine that under some special
 conditions in a finite volume $V$ during the time interval $T$ the
 vacuum condensate may be disoriented in isotopical space.  It is
 convenient to  describe chiral dynamics by  sigma-model
 with isoscalar  $\sigma$ and isovector $\vec{\pi}$ fileds
  (in the case of $N_{f} = 2$)   and   constarint $\sigma^{2} +
 \vec{\pi}^{2} = f_{\pi}^{2}$.  In vacuum one has $<\sigma> =
 f_{\pi}$ and since $\sigma$ is a isoscalar there is an  isoscalar
 condensate $<\bar{\psi}\psi>$  only. However one can consider
 another configuration - $<\sigma> = f_{\pi}\cos\theta$ and
 $\vec{\pi} = f_{\pi} \vec{n} \sin\theta$, here  $\vec{n}$ is some
 unit vector in the isospace,  which describes DCC, i.e.
 some  classical pion field configuration, which is metastable
 and  decays  after some time into pions - the signature
 for this event will be the large number of  either neutral ($\pi^{0}$)
 or charged ($\pi^{\pm}$) pions \cite{dcc}.
 It will be interesting
 to formulate effective low-energy  theory describing this condensate
 and small fluctuations around it.  One obvious candidate is the
 usual four-dimensional chiral model - however we must remember that we
 are considering now the small fluctuations of the order parameter
  not  around
  the $O(3,1)$ invariant
 vacuum, but
   around some  new metastable ground state  arising immediately
 after  the collision where  neither
 $3+1$ Lorentz invariance nor rotational invariance are  valid.
 Thus the effective chiral model may be anisotropic and one can
 think about some new universality classes.

 It is important to remember that in the Bjorken model
 of the hydrodynamical expansion  in the
 central rapidity region  \cite{b}
   (see also \cite{km})
   one has  approximate  $1+1$ Lorentz invariance with respect
 to the longitudinal boosts
   $x^{\pm} \ra \exp(\pm\theta) x^{\pm}$
  which is based on the fact
  that initial conditions
   for the  hydrodynamical  expansion are invariant
 with respect to longitudinal  Lorentz  boosts\footnote{
 This initial conditions are different from the initial
 conditions in  the Landau  model \cite{landau} which leads to
 explosion at $t,z \approx 1 fm$}.    It was
 estimated in \cite{b,km} that until the time $R_{A}/v_{s}$, where
 $R_{A}$ is the nucleus radius and $v_{s}$ is the  speed
 of the sound waves in nuclear matter
 (for a uranium nucleus this time is approximately 10 fm) the
 expansion will be largely $1+1$ dimensional.   It is precisely
 during this time the DCC may be formed and  we shall try to
 find the effectie $1+1$-dimensional theory describing DCC in the
 central rapidity region, where rapidity is  defined   as
  ~~$y = {1\over 2}
 \ln (x^{+} / x^{-})$.

To derive  this  effective  theory we shall  use the same approach
 as in a
  derivation  of a  $3+1$   chiral Lagrangian  from QCD (see, for
 example, \cite{de} and references therein),
 i.e. consider pions $\pi^{a}$ as the elementary
 external fields interacting with  quarks. The effective low-energy
 action  is
\bq
e^{-W[\pi]} & = & Z^{-1} \int DA_{\mu} D\Psi D\bar{\Psi}
 exp \{ -S[A] + \int d^{4}x \bar{\Psi} e^{i\Pi \gamma_{5}}
[\gamma_{\mu}(i\partial_{\mu} + e A_{\mu}) + M] e^{i\Pi \gamma_{5}} \Psi \}
\nonumber \\
Z & =  &  \int DA_{\mu} D\Psi D\bar{\Psi}
 exp \{ -S[A] + \int d^{4}x \bar{\Psi}
[ \gamma_{\mu}(i\partial_{\mu} + e A_{\mu})+M] \Psi \}
\label{effaction}
\eq
where $\Pi = {\pi^{a}t^{a} /  f_{\pi}}$  and $t^{a}$ are the
 generators of the axial $SU(N_{f})$, $f_{\pi} = 95\, MeV$ is
 the pion coupling constant
  and  $S[A]={1\over 4} \int\!d^4x\,
 tr (F_{\mu\nu}F^{\mu\nu})$ is the gluon action,
$ F_{\mu\nu} =
\partial_\mu A_\nu \! - \partial_\nu A_\mu + e \, [A_\mu,A_\nu]$
is the non-abelian field strength,
$A_\mu = A_\mu^a \tau^a$ and
$\tau^a$ are the generators of the  the colour $SU(3)_{c}$. We also
 include  here the quark mass term $\bar{\Psi}M\Psi$, where $M
 = diag(m_{1}\cdots m_{N_{f}})$ is the quark mass matrix.

 We can rewrite the fermion part of the action as:
\bq
S_{\Psi} = \int d^{4}x
 \bar{\Psi}\gamma_{\mu}\large[\frac{1+\gamma_{5}}{2}(i\partial_{\mu}
 +  L_{\mu}) + \frac{1-\gamma_{5}}{2}(i\partial_{\mu} + R_{\mu})\large]
\Psi
\eq
where left and right gauge fields are
\bq
L_{\mu} = I \otimes  eA_{\mu} +  i U^{-1}\partial_{\mu} U
\otimes I \nonumber \\
R_{\mu} =I \otimes  eA_{\mu} +  i U \partial_{\mu} U^{-1}
\otimes I
\label{LR}
\eq
and $U = \exp(i\Pi)$. The first factor in the direct products refers
 to flavor $SU(N_{f})$, while the second refers to color $SU(3)_{c}$.

 We are looking now for some new  (quasi)one-dimensional chiral models
  and will try to derive
  some  effective $1+1$-dimensional action $W_{1+1}[\pi]$ describing
 the fluctuations of the chiral order parameter in the central
  rapidity region arising after nucleus-nucleus collision. To extract
 that part of the action that is relevant to the high-energy collision
   we shall use rescaling of the light-cone coordinates  suggested
 in \cite{verlinde} $x^{\pm} \rightarrow \lambda x^{\pm}$.
The components of the gauge potential are transformed  under rescaling
as $A_i \ra A_i$ and  $A_\alpha \ra \lambda^{-1} A_\alpha$, while the
 quark fields are transformed as $\psi \ra \psi/ \sqrt{\lambda}$.
Hence the rescaled gluon action  can be written in the following form
\bq
\label{rescaling}
S =  \int d^{4}x \; tr[{1\over2} ( E^{\alpha\beta}F_{\alpha\beta} +
 F_{\alpha i}F^{\alpha i}) + {\lambda^2\over 4}
(E_{\alpha\beta}E^{\alpha\beta}+F_{ij}F^{ij})]
\eq
where $E_{\alpha\beta}$ is an auxilary field. The rescaling action
 describes the same physics as the original one if one also rescales
 $s \ra \lambda^{2} s$, thus high energy limit $s \ra \infty$
 corresponds to $\lambda \ra 0$ in the rescaled theory \cite{verlinde} and
 we get the truncated action
\bq
\label{Sred}
S[A,E] ={1\over 2}\int \, d^2 x^{i} dx^{+}dx^{-}
 \, tr(E^{\alpha\beta}F_{\alpha\beta}
+ F_{\alpha i}F^{\alpha i} ).
\eq
and  now  the auxiliary field $E^{\alpha\beta}$  becomes  a Lagrange
multiplier imposing the zero-curvature constraint
\be
F_{\alpha\beta} = 0,~~~~~~ A_{\alpha} = {1\over e} g^{-1}\partial_{\alpha} g
\ee

Now let us consider the fermion  action after rescaling
\bq
S_{\Psi} = \int d^{4}x
 \bar{\Psi}\gamma_{\alpha}\large[\frac{1+\gamma_{5}}{2}(i\partial_{\alpha}
 +  L_{\alpha}) + \frac{1-\gamma_{5}}{2}(i\partial_{\alpha} +
 R_{\alpha})\large] \Psi +
  \nonumber \\
  \lambda \int d^{4}x
 \bar{\Psi}\gamma_{i}\large[\frac{1+\gamma_{5}}{2}(i\partial_{i}
 +  L_{i}) + \frac{1-\gamma_{5}}{2}(i\partial_{i} + R_{i})\large] \Psi
 + \lambda \int d^{4}x  \bar{\Psi}M\Psi
\eq
Taking the limit $\lambda \ra 0$ we  see that mass term disappears and
  quarks propagate
 in the longitudinal directions only and  couple only to $L_{\pm}$ and
 $R_{\pm}$ components which  becomes pure gauge (let's remember
 that colour and flavor commute with each other):
\bq
L_{\pm} = (U\otimes g)^{-1}\partial_{\pm}(U\otimes g),~~~~~~~~~
R_{\pm} = (U^{-1}\otimes g)^{-1}\partial_{\pm}(U^{-1}\otimes g)
\label{LRpm}
\eq

To get the effective two-dimensional action we must rewrite four
 dimensional
 fermions $\Psi$ in terms of two-dimensional ones. Using the chiral basis
 for $\gamma$-matrices:
\bq
\gamma_{0}  = \left(\begin{array}{cc} 0 & I \\
 I & 0 \end{array}\right), ~~~~
\gamma_{i}  = \left(\begin{array}{cc} 0 & \sigma_{i} \\
 -\sigma_{i}  & 0 \end{array}\right), ~~~~
\gamma_{5}  = \left(\begin{array}{cc} I & 0 \\
 0 & -I \end{array}\right)
\eq
 it is easy to get
\bq
S_{1+1}(\psi,\phi) = &\int d^{4}x \,
 \Psi_{L}^{\dagger}\large[(i\partial_{0}+L_{0})-\sigma_{z}
(i\partial_{z}+L_{z})\large]\Psi_{L}  +
  \Phi_{R}^{\dagger}\large[(i\partial_{0}+R_{0})+\sigma_{z}
(i\partial_{z}+R_{z})\large]\Phi_{R} &  = \nonumber \\
&\int d^{2}x^{i} dx^{+}dx^{-} \,
 \psi_{+}^{\dagger}(i\partial_{+} + L_{+})\psi_{+} +
\psi_{-}^{\dagger}(i\partial_{-} + L_{-})\psi_{-} \; + &  \\
&\int d^{2}x^{i} dx^{+}dx^{-} \,
 \phi_{+}^{\dagger}(i\partial_{+} + R_{+}) \phi_{+} +
\phi_{-}^{\dagger}(i\partial_{-} + R_{-}) \phi_{-} \;\;\; &  \nonumber
\eq
where two-component left and right spinors $\Psi_{L}$ and $\Psi_{R}$
  were  defined as $\Psi_{L} = (\psi_{-},\psi_{+})$ and
 $\Psi_{R} = (\phi_{+},\phi_{-})$  and
 there are two left (in two-dimensional sense) $\psi_{+}$ and
 $\phi_{+}$ fermions and two right $\psi_{-}$ and $\phi_{-}$ fermions.

 When we neglected the transverse part of the fermion action we assumed that
 quarks  transverse momenta are sufficiently small $\lambda |p^{\perp}| << 1$
 and it looks reasonable to consider
  fermion fields as independent on $x^{i}$ -
 one has  decoupled one-dimensional systems  and
 to calculate  the effective action $W_{1+1}(\pi)$ we must calculate the
 two-dimensional determinants of the Dirac operators:
   $W_{1+1}(\pi) = \ln det[\sigma_{\alpha}(\partial_{\alpha} + L_{\alpha})]
 + \ln det[\sigma_{\alpha}(\partial_{\alpha} + R_{\alpha})]$.  The determinant
 of the Dirac operator in a nonabelian gauge field $A_{\alpha}$ have been
 calculated by
 Polyakov and Wiegmann \cite{pw}
\bq
\ln det [\sigma_{\alpha}(\partial_{\alpha} + A_{\alpha})] =
 W_{1}(G\cdot H^{-1}); ~~~~
A_{+} = G^{-1}\partial_{+} G, ~~ A_{-} = H^{-1}\partial_{-} H
\label{det}
\eq
 where  $W_{k}(G)$ is the action \cite{nw} of the Wess-Zumino-Novikov-Witten
(WZNW)
 model at level $k$
\bq
 W_{k}(G)={k\over 4\pi}\int d^2 x  Tr\left(
      G^{-1}\partial_{\alpha}G\cdot G^{-1}\partial_{\alpha}G\right) + \nonumber
\\
 {k\over 12\pi}\int_M d^3 x\,\,\epsilon^{\alpha\beta\gamma}
Tr G^{-1}\partial_{\alpha}G\cdot
    G^{-1}\partial_{\beta}G\cdot G^{-1}\partial_{\gamma}G,
\eq
and $M$ is a three-dimensional disc the boundary of which is our
 two-dimensional space. The chiral field $G(x)$ takes the value in
 the direct product $SU(3)_{c} \times SU(N_{f})$.

 Using (\ref{LRpm}) one can see that  for both $L_{\pm}$ and $R_{\pm}$
 fields $G_{L(R)} = H_{L(R)}$ and thus $W_{1+1}(\pi) =  W(G_{L}
\cdot H_{L}^{-1}) + W(G_{R} \cdot H_{R}^{-1})  = 0$, i.e. we
 do not obtain any nontrivial two-dimensional  chiral action -  the chiral
dynamics
 of DCC  will be described by usual four-dimensional chiral lagrangian.
 The reason for this is the follows -  after the reduction
 to a two-dimensional problem the chiral (in a four-dimensional sense)
 spinors $\Psi_{L}$ and $\Psi_{R}$ were transformed into  left-right
 symmetric (in a two-dimensional sense) pairs $\psi_{\pm}$ and
 $\phi_{\pm}$ interacting with two vector fields $L_{\pm}$ and $R_{\pm}$
  - thus in  resulting theory original chiral rotations
 look like  vector gauge transformations and have no anomaly, which
 means that  there is no two-dimensional  chiral
 action . To obtain nontrivial
 $W_{1+1}(\pi)$ one must have two-dimensional   anomaly for chiral
transformation
 and it is possible to obtain it by  removing left(right)-moving fermion
 from one pair and right(left)-moving fermion from another one
 \footnote{We cannnot remove both $\psi_{+}, \phi_{+}$ or
 $\psi_{-}, \phi_{-}$ because in this case there  will be anomaly
 for color $SU(3)_{c}$ too,  which is absolutely forbidden}.
  What means this further reduction ? Let us
 remember that chiral fermions have definite  spin projection
   on the momentum direction $\vec{n}$ (neglecting the fermion mass)
    $(\vec{\sigma}\vec{n})\Psi_{L,R} = \pm \Psi_{L,R}$.  Then afer
 reduction one has four $1+1$-dimensional fermion moving along $z$-axis -
$\psi_{\pm}$ and $\phi_{\pm}$. For $\psi$ fermions (which are components of
 left spinor $\Psi_{L}$)  the direction of spin
 is the same as the direction of momentum, for $\phi$ it is the opposite.
 Thus removing, let us say, $\psi_{-}$ and $\phi_{+}$ we have
 one left-mover $\psi_{+}$ and one right-mover $\phi_{-}$ - but  both of
 them have the {\it same} direction of spin and  this
 additional reduction leads to the spin-polarized  vacuum. One can estimate
 the spin in this state as a volume in a phase space occupied by
 this fermions $S \approx R_{\perp}^{2} L \Lambda^{3}$, where
 $R_{\perp}$ is the transverse scale, $L$ is the longitudinal one and
 $\Lambda$ is the ultraviolett cutoff for effective chiral model
 which is defined by the inverse "size of the pion", i.e. pion
 coupling constant  $f_{\pi} = 95 MeV$. One can estimate transverse
 radius  as  $R_{\perp} \sim 1/f_{\pi}$  comparing the two-dimensional
reduction
  of the
 four-dimensional action $f_{\pi}^{2}\int d^{4}x \partial_{\mu}\pi
 \partial_{\mu}\pi$ and the two-dimensional one
$ \int d^{2} x \partial_{\alpha}\pi  \partial_{\alpha}\pi$.
 The fact that $R_{\perp}f_{\pi}\sim 1$ means that  we  really
 have one-dimensional phase space (as it should be) and transverse
 fermion excitations  are irrelevant. For spin one gets estimate
  $S \approx L f_{\pi}$, for example  for $L = 10$ fm. one gets $S \approx 10$.

 We can suggest that such a  spin-polarized state  can appear
  in ultra- relativistic collisions   with   polarized nucleus.
  To support this idea let us use the Walker  arguments \cite{walker}
 about the heavy ion as a color vacuum-cleaner - due to the strong
 interaction of each colored parton  in the target
   with soft gluons from  projectile all colored degrees of freedom will
 be swept out of  the target. That "nothing" which  is left behind the
 leading particles in the central rapidity region  carries no
 memory of the valence degrees of freedom, but carries four-momentum
 $P_{\mu}$ and it is in this region the creation of DCC is possible.
  However the same
  region   can carry angular momentum too - and thus one can imagine
 that  after the collision of the polarised nucleus there will be
 large spin polarization in the central rapidity region. Then one
 can think that  reduced  $1+1$-dimensional model with action
(again we   are assuming that there are no dependence on the
 transverse coordinate $x^{i}$)
\bq
S_{1+1}(\psi_{+},\phi_{-}) =
 \int d^{2}x^{i} dx^{+}dx^{-}\large{
 \psi_{+}^{\dagger}(i\partial_{+} + L_{+})\psi_{+} +
 \phi_{-}^{\dagger}(i\partial_{-} + R_{-}) \phi_{-} \large}
\label{psiphi}
\eq
will describe spin-polarized DCC.
 The effective action $W_{1+1}(\pi)
 = \ln  \det [\sigma_{\alpha}(\partial_{\alpha} + A_{\alpha})]$
is nontrivial now, because the gauge fields are
\bq
A_{+} = (U\otimes g)^{-1}\partial_{+}(U\otimes g),~~~~~~~~~
A_{-} = (U^{-1}\otimes g)^{-1}\partial_{-}(U^{-1}\otimes g)
\eq
so $G = U\otimes g,~~H = U^{-1}\otimes g$ and $G\cdot H^{-1} = U^{2}
\otimes I$. The gluon degrees of freedom are decoupled from the
 effective action  and trace over the color group gives us the factor
 $N_{c} = 3$ in front of WZNW  action.
  Thus the effective one-dimensional chiral dynamics of the
 spin-polarized disoriented chiral condensate is governed by
 the  $SU(N_{f})$  WZNW  model at level $k = 3$:
\bq
W_{1+1}[\pi] =
 {3\over 4\pi}\int d^2 x  Tr\left(
      U^{-1}\partial_{\alpha} U\cdot U^{-1}\partial_{\alpha}U\right)
 + \nonumber \\
 {3\over 12\pi}\int_M d^3 x\,\,\epsilon^{\alpha\beta\gamma}
Tr U^{-1}\partial_{\alpha}U\cdot
    U^{-1}\partial_{\beta}U\cdot U^{-1}\partial_{\gamma}U,
\eq
where we changed definition and denote   $U^{2}$ as $U$,
 which is now defined as  $U = \exp(2i\pi^{a} t^{a} / f_{\pi})$.
  May be it is also possible to get WZNW models at levels $2$ and $1$ -
 in the case when the spin-polarization is not complete and besides
 $\psi_{+}$ and $\phi_{-}$ fields there are some (but smaller)
  number of $\psi_{-},\phi_{+}$ pairs - however this  is not completely
 clear now.

Thus
 long-range fluctuations of the order parameter $U$
 are described by the two-dimensional  conformal field theory -
 this is important and means that one can have quasi one-dimensional chiral
 condensate which will not be destroyed by the infrared effects.
  The WZNW model is exactly solvable and the spectrum of
 anomalous dimensions and correlation functions are known \cite{kz},
 so we can use this information to study the correlation between
 th order parameter values in the
 different  space-time  regions.
 The two-point correlation function (up to some  normalization
 factor $C$) is
\bq
 \langle U^{i}_{j}(z,t) ~U^{\dagger}{}^{k}_{l} (z',t') \rangle  =
C \delta^{i}_{l}\delta^{k}_{j} [(z-z')^2 - (t-t')^2]^{-2\Delta}
\label{2point}
\eq
  $i,j,k,l$ are $SU(N_{f})$ indices and anomalous dimension
 of the  chiral field $U$  equals
\bq
\Delta = \frac{ c_{g}}{N_{f} + k}
\eq
where the constant $c_{g}$ is defined as $t^{a}t^{a} = c_{g} I$.
For flavor $SU(2)$ and $k=3$  one gets $c_{g} = 3/4$ and  $\Delta = 3/20$.
It is more convenient to rewrite  correlation function (\ref{2point})
 in proper time - rapidity coordinates \cite{b}
$\tau^{2} = t^{2} - z^2;~ y = (1/2) \ln (t+z)/(t-z)$ and to study
 the correlation at equal proper time $\tau$.  Then
\bq
 \langle U^{i}_{j}(\tau,y) ~U^{\dagger}{}^{k}_{l} (\tau, y') \rangle  =
 C \delta^{i}_{l}\delta^{k}_{j}(\sqrt{2} \tau)^{-4\Delta}
 [ch(y-y')-1]^{-2\Delta}
\label{2point'}
\eq
and in the central rapidity region when $y$ and $y'$ are small
 we shall get  a simple scaling law $(y-y')^{-3/5}$.
 Because the direction of the order parameter gives us the
 ratio   of neutral to charge pions arising after decay of DCC,
 the two-point correlation function  (\ref{2point'})
 gives us a  distribution  of the
 events with a given ratio of neutral to charged pions  in a
 given rapidity interval. We also can consider more complicated
 correlation functions, for example four-point correlation functions
 which were calculated in \cite{kz}. These   function will give us
 the multiparticle correlations. It will be extremely
  interesting to look for such correlation either in nucleus-nucleus
 collisions or even at high-energy nucleon-nucleon collision at SSC.
It is also interesting to understand  what will be the phenomenological
 manifestations of the spin-polarization.

In conclusion we would like to  briefly discuss the possibility to
 take into account the effect of leading particles (the regions
 in the rapidity space with the nonvanishing baryon number) in
 this picture. What we have is the two-dimensional conformal field theory
 defined on the finite  interval. The baryon charge is distributed
 at the boundaries of this interval and one can try to
 take it into account by adding some boundary vertex operators
 - if this is possible one can apply the well eleborated methods
 of the conformal field theory  to study the dynamics of the
 condensate at large rapidities. The  effect of the baryon charge
  in this regions could be described as the soliton charge of
  the   boundary vertex operators. Using this approach
 one can  study  correlations between leading particles
 moving into opposite directions.   It is also interesting to understand
 how to take into account  interaction
 between fluctuations of the DCC order parameter $U$ and  the
 hydrodynamical fluctuations during the $1+1$-dimensional  expansion.
 The hydrodynamical fluctuations  are connected with the
   group of the  one-dimensional diffeomorphisms  $Diff R^{1}$
   and  it may be that the large scale fluctuations are governed
 by the two-dimensional quantum gravity. These  and related questions
 deserve more detailed investigations.

I am grateful to A.A. Anselm for introducing me into this subject
 and to A. Bilal, I. Klebanov and A. Polyakov for interesting and
 stimulating discussions. This work was supported by the National
 Science Foundation grant NSF PHY90-21984.

Note added. - While this paper had  being prepared for publication I was
 aware about recent paper of Khlebnikov \cite{khlebnikov} where
 the two-dimensional description of the DCC was considered and WZNW
 model was  suggested as infrared fixed point of the general
 asymmetric chiral model. It was also suggested that one
 can get WZNW model from reduction of the $3+1$ model with the
 Wess-Zumino term  provided there is topologically  nontrivial configurations
 of kaons in transverse directions.   However
  the level of the WZWN model in \cite{khlebnikov}
  was suggested to be $1$, not $3$. which leads
 to non-integer topological charges of the kaon configuration $Q = 4/3$.
  Let us note,
 that if one takes $k=3$ instead of $k=1$  the charge will be integer $Q=1$.
 The relation between this reduction and the spin-polarization picture
 suggested in this paper is unclear for us now.


\begin{thebibliography}{99}
\bibitem{lipatov}
L.N. Lipatov, Nucl. Phys. {\bf B365} (1991) 614
 and review  in {\it Perturbative QCD}, ed. A.H. Mueller
(World Scientific, Singapore, 1989), and references therein;
\bibitem{verlinde} H. Verlinde and E.Verlinde, Princeton preprint
 PUPT-1319, IASSNS-HEP-92/30 , hep-th 9302104  (1993).
\bibitem{dcc}
A.A. Anselm,  Phys. Lett. {\bf B217} (1989) 169;\\
A. A. Anselm and M. G. Ryskin, Phys. Lett. {\bf B266}  (1991) 482;\\
J. D. Bjorken, Int. J. Mod. Phys. {\bf A7} (1992) 4189  and
Acta Physica Polonica {\bf B23}  (1992) 561; \\
J.-P. Blaizot and A. Krzywicki, Phys. Rev. {\bf D46}  (1992) 246; \\
K. L. Kowalski and C. C. Taylor, Case Western Reserve University preprint
92-6, hep-ph/9211282 (1992);\\
K. Rajagopal and F. Wilczek, Princeton preprints PUPT-1347,
IASSNS-HEP-92/60, hep-ph/9210253 (1992) and PUPT-1389, IASSNS-HEP-93/16,
hep-ph/9303281 (1993).
\bibitem{b}
J. D. Bjorken, Phys. Rev. {\bf D27}  (1983) 140.
\bibitem{km}
K. Kajantie and L. McLerran, Nucl. Phys. {\bf B214}  (1983) 261.
\bibitem{landau}
L.D. Landau, Proc.Acad.Sciences USSR, Physical Series {\bf 17} (1953) 51
 (in Russian).
\bibitem{de} D.I. Diakonov and M.I.Eides, JETP. Lett. {\bf 38} 433.
\bibitem{pw}
A. M. Polyakov and P. B. Wiegmann, Phys. Lett. {\bf 131B} (1983) 121,
 {\bf 141 B} (1984) 223.
\bibitem{nw}
S. P. Novikov, Usp. Mat. Nauk, {\bf 37}  (1982) 3; \\
E. Witten, Comm. Math. Phys. {\bf 92}  (1984)  455.
\bibitem{walker} W.D. Walker, March 1992, unpublished,
 quoted in J. D. Bjorken,
Acta Physica Polonica {\bf B23}  (1992) 561.
\bibitem{kz}
V. G. Knizhnik and A. B. Zamolodchikov, Nucl. Phys. {\bf B247}
(1984) 83.
\bibitem{khlebnikov} S.Yu. Khlebnikov, preprint UCLA/93/TEP/10,
 hep-ph/9305207
\end{thebibliography}
\end{document}